**Tailoring the molecular structure to suppress extrinsic disorder in organic transistors**

*Nikolas A. Minder,\* Shaofeng Lu, Simone Fratini, Sergio Ciuchi, Antonio Facchetti, and Alberto F. Morpurgo\**

[*]     N. A. Minder, Prof. A. F. Morpurgo
DPMC and GAP, University of Geneva
24 quai Ernest-Ansermet,
CH - 1211 Geneva (Switzerland)
E-mail: nikolas.minder@unige.ch, alberto.morpurgo@unige.ch
        Dr. Shaofeng Lu, Prof. A. Facchetti
Polyera Corporation, 8045 Lamon Avenue
Skokie, IL 60077 (USA)
        Dr. Simone Fratini
Institut Néel - CNRS & Université Joseph Fourier,
BP 166, F-38042 Grenoble Cedex 9 (France)
        Prof. Sergio Ciuchi
Istituto dei Sistemi Complessi CNR, CNISM and Dipartimento di Fisica,
Università dell'Aquila, via Vetoio,
I-67100 Coppito-L'Aquila (Italy)

In the absence of extrinsic disorder, the mobility of charge carriers in top-quality organic field-effect transistors (FETs) is expected to increase upon lowering the temperature $T$.[1-10] Such a behavior is analogous to that observed in inorganic semiconductors, exhibiting conventional band transport, where charge carriers are described by wave packets undergoing rare scattering events. In organic semiconductors based on small conjugated molecules, however, the underlying physics is different and poorly understood microscopically.[11, 12] Even in the best single-crystals, the estimated carrier mean free path at room temperature is comparable to the lattice spacing,[10, 13] which prevents the use of the band picture. Indeed, thermal molecular vibrations induce very large fluctuations in the inter-molecular hopping integrals[5, 14] that – on the much faster time scales characteristic of electronic motion – are experienced as strong static disorder, and cause Anderson localization.[15] On a longer time scale, the localized carriers can still propagate, driven by the molecular dynamics.[5, 10, 16] In this scenario (referred to as "band-like transport"[17, 18]), the mobility increases as $T$ is lowered, because the amplitude of the molecular motion – and hence the hopping integral fluctuations – are smaller at lower $T$, resulting in a longer localization length.[7, 13, 16] At very low temperature, when the localization length eventually becomes much longer than the lattice spacing, true band transport is expected.[9, 19] However, this regime is not reached in the experiments even in organic FETs based on the highest quality molecular single crystals: as $T$ is lowered below a temperature $T^*$ (that is material and device dependent), extrinsic sources of disorder contribute to localize carriers, causing a drastic suppression of their mobility.[17, 20-22]

To access the intrinsic transport properties of organic FETs in a broad temperature range – as it is essential to gain a true microscopic understanding of the band-like transport regime – materials and devices are needed, which are less affected by extrinsic disorder. This is difficult, because it is not yet established what the dominant sources of disorder in high-quality organic single-crystal FETs are. Nevertheless, useful information has been obtained from comparative studies of organic FETs with different gate dielectrics, which have shown how the coupling of charge carriers to their polarizable environment strongly enhances the tendency towards carrier localization.[23-28] Different microscopic mechanisms may play a role, such as the formation of interfacial polarons,[25, 26] electrostatic fluctuations generated by dipolar disorder,[23, 28] interplay between disorder and the effect of electrical polarizability,[24, 27] etc. Despite the differences, all of these mechanisms share a common aspect: disorder and the tendency towards localization originate from – or are enhanced by – the electrostatic interaction between a charge carrier in the channel and nearby charges, either unintentionally present, or due to polarization effects (e.g., image charges in the dielectric or in the semiconductor itself).

Based on this concept, we have recently proposed structure-property relationships which are expected to favor the occurrence of band-like transport in high mobility molecular semiconductors.[21] The most transparent of these relations holds for organic crystals of molecules consisting of identical conjugated cores (that, in a crystal, are π-stacked to form the conducting crystalline planes), functionalized with chains (that determine the spatial separation between adjacent conducting crystalline planes). Our prediction is that the molecules with the longest substituents lead to a more pronounced band-like transport, with the carrier mobility increasing down to a lower temperature $T^*$ (a conclusion that holds under the assumption that changing the ligands does not significantly modify the crystal packing of the conjugated molecular cores). Indeed, longer core substituents increase the spatial separation between the charge carriers in the FET channel – that reside on the conjugated cores – and other charges in the system that contribute to enhance extrinsic disorder. Since all electrostatic interactions decrease with increasing distance, a larger spatial separation results in a weaker electrostatic coupling and, hence, smaller disorder strength.

Here, we validate this idea with a set of transport experiments (and their quantitative analysis) on FETs realized on single crystals of two perylene derivatives, PDI3F and PDI5F (**Figure 1**). These core-cyanated molecules [$N,N'$-bis(R)-(1,7 and 1,6)-dicyanoperylene-3,4: 9,10-bis(dicarboximide)] have the same conjugated core but two different substituents on the nitrogen atoms: R = $CH_2C_3F_7$ for PDI3F (also known as PDIF-$CN_2$[29, 30]); R = $CH_2C_5F_{11}$ for PDI5F. We find that in the temperature range where the mobility increases upon lowering $T$ – i.e., when transport is determined by the intrinsic properties of the materials – FETs of both molecules exhibit virtually identical mobility values. In the devices based on the molecule with shorter side ligands, however, the mobility starts decreasing at a higher temperature. The effect is very reproducible, and the analysis of the full temperature dependence of the mobility is quantitatively consistent with the original idea, namely that the longer ligand length contributes to suppress extrinsic disorder generated by charges located outside the FET channel. These results provide important microscopic insights on the effects of the molecular and crystal structure on the performance of organic FETs. They also give information about the microscopic origin of extrinsic disorder limiting transport in high-quality organic transistors.

Although the underlying idea – comparing the temperature dependent mobility in transistors based on two different, but similar, molecules – is straightforward, the experiments are challenging, because of the very stringent conditions that they impose on the molecules to be used. First – and obviously – devices need to be realized in which the band-like transport regime is experimentally visible. Many different molecules with the structural motif of non-conjugated ligands of variable length attached to π-conjugated molecular cores have been used to realize organic FETs, such as $C_n$-BTBT,[31] $C_n$-DNTT,[32] $C_n$-BDT,[33] and PTCDI-$C_n$,[34] and band-like transport in single-crystal FETs has been reported in rare cases.[35, 36] However, experimental reproducibility is an issue:[37] to detect the change in behavior induced by a small change in the length of the molecular ligands, highly reproducible measurements on many devices are needed, to ensure that differences are not just originating from small statistics and sample-to-sample fluctuations.

Band-like transport together with a high level of experimental reproducibility for single-crystal FETs of organic semiconductors has been achieved only with a few materials,[17, 20-22, 38] always using crystals grown by physical vapor phase transport (VPT). Among these materials, core-cyanated perylenes are an excellent choice for this study as functionalization of the nitrogen atoms with various substituents could access a library of derivatives. We therefore started by synthesizing four PDIR-CN$_2$ derivatives having R = CH$_2$CF$_3$ (PDI1F), R = CH$_2$C$_3$F$_7$ (PDI3F), R = CH$_2$C$_5$F$_{11}$ (PDI5F), R = CH$_2$C$_7$F$_{15}$ (PDI7F). In the case of PDI1F and PDI7F, purification of the synthesized molecules proved to be extremely difficult. Only small amounts of these molecules could be produced with sub-optimal purity, and single crystals grown by VPT were irregularly shaped and showed poor, irreproducible FET characteristics (for these molecules, the same problems had been found previously in thin film devices[39]). For PDI5F, on the contrary, sufficiently large quantities of high-purity materials could be synthesized, enabling the growth of well-shaped crystals leading to highly reproducible, high-mobility OFETs (see below).

Having identified two molecules with the desired structural motif that enable the reproducible observation of band-like transport in single-crystal OFETs, we have checked that the molecular core packing in bulk crystals is approximately the same in the two cases. Indeed, it often happens that drastic differences in the structure of molecular crystals – leading to large changes in orbital overlaps and in their fluctuations, i.e. precisely those parameters which govern charge transport – are caused by seemingly small changes in the constituent units.[40] By means of thorough X-ray diffraction measurements, we have resolved the crystalline structure of the two materials (see Figure 1 and Experimental Section), from which we conclude that the structural differences between PDI3F and PDI5F crystals do not lead to very significant changes in the relevant transport parameters. Variations in the transfer integrals and in their dynamical fluctuations could arise from differences in the inter-molecular separation, the translational displacements of the perylene cores of adjacent molecules,[41, 42] and in relative orientation. However, in PDI3F and PDI5F crystals, the minimum π-π stacking distance $d(π-π)_{min}$ differs by less than 0.5 % (3.358 Å in PDI3F and 3.343 Å in PDI5F). As for differences in translational displacement of perylene cores in the two crystals, this is at most 0.2 Å, much smaller than the size of a benzene ring. Finally the α, β, and γ angles defining the unit cell are the same for PDI3F and PDI5F within approximately 3 %. Therefore, also on the basis of the crystalline structures, the two molecules appear to be ideally suited to investigate the influence of the ligand length on band-like transport in organic semiconductors.

We have realized many FETs of PDI3F and PDI5F by laminating thin VPT-grown single-crystals on so-called air-gap PDMS stamps[43] (see Figure 1d and 1h), Cytop, and PMMA gate dielectrics, using techniques that are by now well-established and characterized[44] (see Supporting Information). All these devices show virtually ideal electrical characteristics (see **Figure 2**a,b for PDMS stamps), no hysteresis and an essentially gate voltage-independent carrier mobility $\mu = \frac{1}{C_i}\frac{d\sigma}{dV_{gs}}$ (see Figure 2c,d for devices on PDMS stamps; $C_i$ is the capacitance per unit area and σ is the $V_{gs}$-dependent electrical conductivity measured in a four-terminal configuration – to eliminate possible contact effects). The data also show that the reproducibility of the temperature dependence of the mobility in different devices is excellent for both molecules (see Figure 2e,f for PDMS stamps and Supporting Information for devices on Cytop and PMMA), which allows us to reliably compare and analyze the *μ(T)* curves measured on the two materials. We first discuss the qualitative aspects of the data and then move on to a more quantitative analysis.

**Figure 3**a-c show the experimentally determined *μ(T)* for PDI3F (full squares) and for PDI5F (open circles), with vacuum (a), Cytop (b), and PMMA (c) gate dielectric (in each case the data are obtained by averaging measurements on approximately five different, but nominally identical, devices). The mobility decreases with increasing the dielectric constant of the gate insulator, as expected, and in all cases band-like transport remains visible in part of the temperature range investigated. Noticeably, Figure 3a shows that throughout the temperature interval where *μ* increases with decreasing *T* the mobility measured on single-crystal FETs of the two molecules essentially coincide. This observation shows that in the regime where the experiments probe the intrinsic material properties, the two materials do behave identically, consistently with the nearly identical packing of the conjugated molecular cores in the two crystals. Deviations in the behavior of the two molecular materials, however, become increasingly pronounced in the range where *μ* decreases upon decreasing *T*, i.e., when the transport properties are determined by extrinsic disorder.[10] Specifically, we observe that for PDI5F, the temperature $T^*$ at which *μ* is maximum is always lower than for PDI3F (for air-gap FET, by as much as ~40 K), and that at low temperature, the mobility is higher than for PDI3F. Indeed, this is the behavior that we had anticipated: the disorder experienced by charge carriers is different in crystals with constituent molecules having different chain lengths, and is effectively smaller for longer substituents.

For a more quantitative analysis we fit the experimental data using a phenomenological mobility edge model that we have developed in the past to describe band-like transport in organic single-crystal FETs,[20] and that has been thoroughly tested.[21, 22] The model describes (see Supporting Information) the electronic properties of the organic materials in terms of a band of delocalized states, whose intrinsic mobility is given by $\mu(T) = \alpha \cdot T^{-2}$ [5, 10] and by a Gaussian tail of states at energies below the band edge, in which carriers are assumed to be fully localized and not to contribute to transport. The device-dependent fitting parameters are the density of states in the tail $N_t$, and the width of the Gaussian tail $E_t$, which is a measure of the strength of disorder ($E_t = 0$ corresponds to having no localized states, and hence no disorder). The value of $\alpha$ is fixed once and for all, and taken to be the same ($\alpha = 3.5 \cdot 10^5$ cm$^2 \cdot$K$^2$/V$\cdot$s) for all devices analyzed – approximately 30 – irrespective of the molecule and of the gate dielectric, since it describes an intrinsic property of the crystalline planes formed by the perylene cores that are the same in all cases (the density of states in the band also enters the model, and is assumed to be the constant, equal to the density of molecules at the surface, divided by the electronic bandwidth ~0.5 eV). Analyzing the data in this way is crucial to properly take into account experimental factors that have a strong influence on the $\mu(T)$ curve, such as the carrier density at which the measurements are performed (the carrier density is different in air-gap as compared to Cytop and PMMA devices, owing to the different breakdown field and dielectric constants of the different device structures).

The model fits well nearly all the $\mu(T)$ curves measured on individual transistors, with the band-tail width $E_t$ playing the most crucial role in determining the device behavior. The excellent agreement with the data is shown in Figure 3a-c, where the continuous lines represent the best fits to the averaged $\mu(T)$ curves measured for PDI3F and for PDI5F air-gap, Cytop, and PMMA transistors. The results of the data analysis are illustrated in **Figure 4**. Figure 4a shows the temperature $T^*$ at which the mobility is maximum as a function of the value of $E_t$ extracted by fitting the $\mu(T)$ curve, for each individual device. Clearly, $T^*$ is smaller for smaller $E_t$, as expected: band-like transport is seen down to lower temperature in devices where the width of the band tail – hence the strength of disorder – is smaller. Much more importantly, Figure 4a shows that the values of $E_t$ extracted from any PDI5F device is always smaller than the value of $E_t$ obtained from PDI3F devices on the same dielectric (see also Figure 4b for the $T^*$ vs. $E_t$ relation averaged over different devices). This systematic behavior is summarized in Figure 4c, which shows the average values of $E_t$ for the different dielectrics, for both PDI3F (full squares) and PDI5F (empty circles). Finding such a systematic behavior excludes that sample-to-sample variations are responsible for our observations. We therefore conclude that the chain length does have a clear effect on the disorder strength (quantified by $E_t$), and influences the tendency of the material to show band-like transport (as quantified by $T^*$). Indeed, extending the chains in PDI3F by two $C_2F_4$ units systematically lowers $T^*$ by approximately 30-40 K in air-gap devices.

It is also apparent from Figure 4b that, at a quantitative level, changing the side chain length has an effect comparable to a change in the gate dielectric: the value of $E_t$ for PDI3F on air-gap devices is approximately the same as that in PDI5F FETs on Cytop, and for PDI3F devices on Cytop, $E_t$ nearly coincides with the value for PDI5F on PMMA. Finding systematically comparable magnitudes of $E_t$ suggests a common origin of disorder in all cases. For FETs with polymeric gate dielectrics, such as PMMA and Cytop, disorder is largely due to potential fluctuations generated by permanent, randomly oriented electrical dipoles that are present on the monomer units of the polymeric dielectric (the so-called dipolar disorder[23, 28]). We attribute also the disorder measured on air-gap devices to potential fluctuations generated by charges adsorbed at the crystal surface (see Figure 4d,e for a schematic representation). We can then check whether the order of magnitude of $E_t$ in all different cases (different dielectrics and different substituent lengths) is consistent with this hypothesis.

To this end, we first determined theoretically the potential fluctuations induced by a random distribution of charged impurities placed on the surface of the organic crystal, at a distance $d$ from the molecular cores (for air-gap devices, charges present in adsorbates at the surface are the dominating source of potential fluctuations). Following Ref. [45], the resulting electrical potential in the conducting channel is statistically distributed with an energy spread $\Delta_{ch}$

$$\Delta_{ch}^2 = 2\pi\, n_{imp}(Q^2/\tilde{\varepsilon})^2[E_i(2q_s d)e^{2q_s d}(1 + 2q_s d) - 1], \qquad (1)$$

which depends crucially on the channel-impurity distance $d$. Here $Q$ and $n_{imp}$ are the charge and density of impurities, $\tilde{\varepsilon} = (\varepsilon + \varepsilon_r)/2$ the effective dielectric constant of the interface, with $\varepsilon$ and $\varepsilon_r$ referring to the organic crystal and to the gate insulator, $E_i(x)$ is the exponential integral function and $q_s$ the inverse screening length characterizing the screening by mobile charges in the channel (see Supporting Information). Eq. 1 shows that $\Delta_{ch}$ is a decreasing function of the distance $d$, which is proportional to the chain length. Longer substituents therefore result in a reduced amount of extrinsic disorder.

For a quantitative comparison of different devices, we use the same impurity density $n_{imp}$ in all cases, whose value we determine experimentally by assuming that the width $E_t$ of the states in the band tail of our phenomenological model is proportional to the spread of potential fluctuations $\Delta_{ch}$. Moreover, we identify $\Delta_{ch} = E_t$, as any proportionality factor would have the only effect of changing the estimated value of $n_{imp}$, without affecting the comparison of different devices. Taking $Q=e$, the distance $d$ equal to half the interlayer spacing $c$ in the molecular crystal (see Figure 4d), the inverse screening length $q_s = 0.02$ Å$^{-1}$ as a representative value in the explored temperature and density range, and $\varepsilon = 3$ we obtain $n_{imp}$ = 1.1*10$^{11}$ cm$^{-2}$ from the experimental value $E_t$ = 33.0 meV for PDI5F/air-gap devices. Using the same parameters and the length $d$ appropriate to PDI3F/air-gap devices yields $E_t$ = 36 meV, in good agreement with the experimental value $E_t$ = 38.2±1.5 meV.

Following the same approach, we include dipolar disorder in devices with polymer gate insulators. It can be shown (it will be discussed in detail elsewhere) that the potential fluctuations originating from randomly oriented dipoles uniformly distributed in a polymer placed at a distance $d$ from the conducting channel have a spread

$$\Delta_{dip} = \sqrt{\frac{\pi}{3}\frac{n_{dip}}{d}}\frac{e\delta}{\tilde{\varepsilon}}, \quad (2)$$

where $\delta$ is the elementary dipole in a monomer unit, $n_{dip}$ is the density of monomers and $e$ the electron charge (since potentials from randomly oriented dipoles are already short ranged,[28] screening by mobile charges in the channel can be neglected). As expected, dipolar disorder is also a decreasing function of the distance $d$. For a comparison with experiments, the case of PMMA dielectric is particularly useful, because the values of $\delta=1.97D$ and $n_{dip} = 7.1\cdot10^{21}$ cm$^{-3}$ are known from the literature,[28] yielding $\Delta_{dip}$ = 44 meV for the PDI5F/PMMA device, and $\Delta_{dip}$ = 48 meV for PDI3F/PMMA. The potential fluctuations generated by the dipoles in the dielectric add to those due to charges adsorbed at the crystal surfaces that we obtained from the analysis of air-gap devices (where now the appropriate dielectric constant has to be used in Equation 1), resulting in a total spread $\Delta_{tot} = \sqrt{\Delta_{ch}^2 + \Delta_{dip}^2}$ of potential values. The resulting values are $\Delta_{tot}$ = 48 meV and $\Delta_{tot}$ = 53 meV respectively, in very good agreement with the experimental values $E_t$ = 47.7±0.8 meV and $E_t$ = 57.5±1.6 meV. These considerations indicate that the estimated strength of the disorder experienced by the charge carriers is consistent with electrostatic potential fluctuations generated by charges present outside the FET channel, which we conclude to be the dominant sources of disorder. Consistently with this conclusion, the strength of disorder decreases upon increasing the substituent length and the magnitude of the effect matches quite precisely our theoretical estimates.

These findings validate our initial assumption on the role of the core substituents in facilitating the experimental observation of band-like transport. They also identify electrostatic potential fluctuations as the dominant source of extrinsic disorder in high-quality organic single-crystal field-effect transistors. We want to emphasize the "universality" of these electrostatic interactions, which implies that the physical mechanisms that we have discussed here are insensitive to details and are therefore relevant more in general (i.e., not only in high-quality single-crystal devices). For instance, it has been repeatedly reported in the past that the deposition of self-assembled organic monolayers on $SiO_2$ substrates used for the realization of organic thin-film transistors resulted in improvements of the carrier mobility.[32, 46, 47] We believe that the results reported here provide an explanation of the effect (which had often been tentatively attributed to an improved morphology of thin films deposited onto substrates coated with monolayers), since the role of the self-assembled monolayers is to space away the charges unavoidably present in the $SiO_2$ gate dielectric from the FET conducting channel. This is just an example, and we anticipate that owing to the universality of the phenomenon, the mechanisms illustrated here will play a role in many other molecular systems with the same structural motif.

**Experimental Section**

Due to their extensive length, details on the synthesis of the molecules, on crystal growth, device fabrication, electrical measurements and on the fitting of the *μ(T)* data, as well as a comprehensive discussion on the theoretical analysis are presented in the Supporting Information. The crystal structures of the two molecular crystals relevant for this study are presented below:

*PDI3F ($C_{34}H_{10}F_{14}N_4O_4$):*

$M_r$=804.46; triclinic, P-1; a=5.2320(14), b=7.638(2), c=18.819(5) Å; α=92.512(5), β=95.247(5), γ=104.730(4)°; V=722.522 Å$^3$;

*PDI5F ($C_{38}H_{10}F_{22}N_4O_4$):*

$M_r$=1004.50; triclinic, P-1; a=5.2983(4), b=7.2482(5), c=23.5708(17) Å; α=95.293(6), β=94.274(6), γ=105.082(5)°; V=865.704 Å$^3$.

**Acknowledgements**
We thank A. Ferreira and I. Gutiérrez Lezama for technical assistance. We also gratefully acknowledge financial support from the Swiss National Science Foundation, the center of excellence MaNEP, and NEDO.

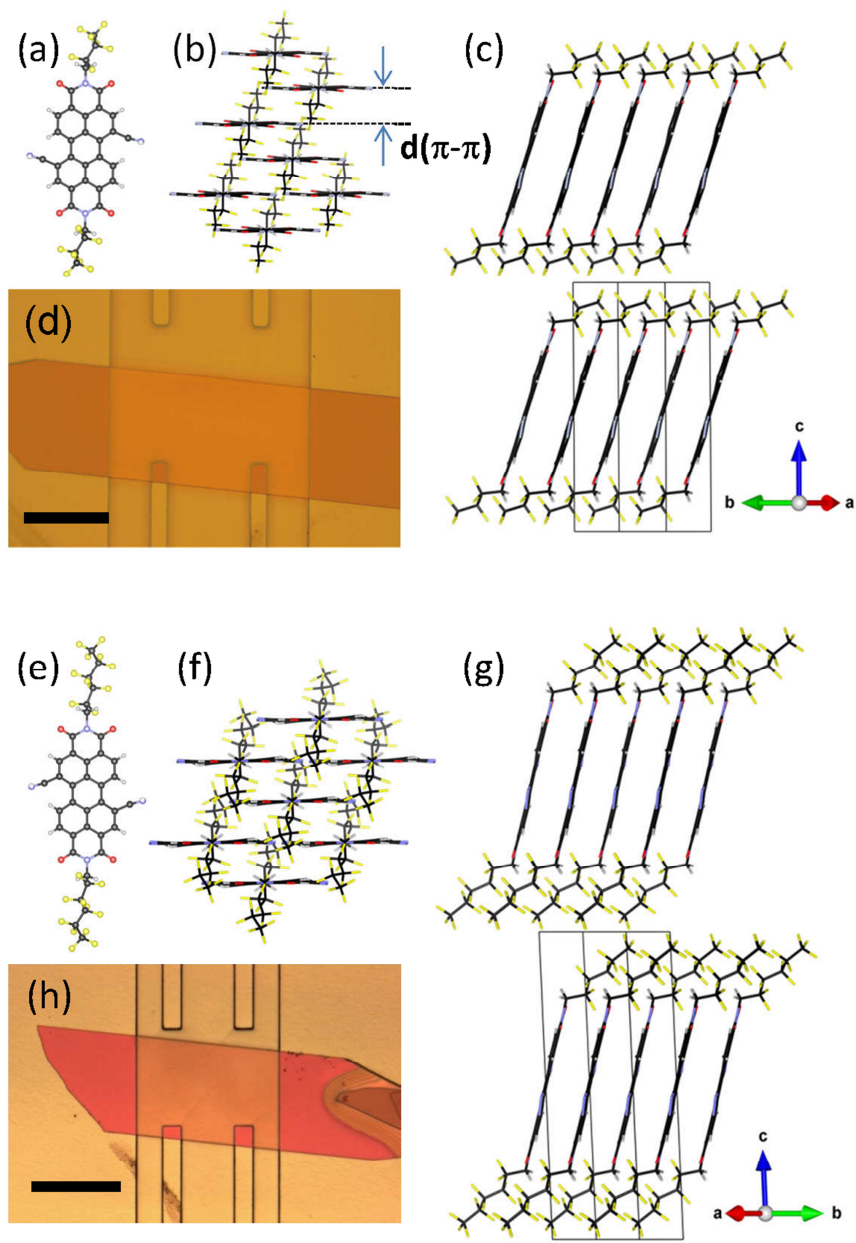

**Figure 1.** Chemical structure and molecular packing of PDI3F (a,b,c) and PDI5F (e,f,g). Panels (b) and (f) show the minimum π-π stacking distance that is the same (within 0.5 %) in the two materials. Panels (c) and (g) illustrate how the separation of the conducting planes formed by the conjugated molecular cores is determined by the chain length. Panels (d) and (h) show optical microscope images of single crystals of PDI3F and PDI5F molecules on PDMS air-gap stamps with a 10 µm recessed gate; the scale bar is 100 µm.

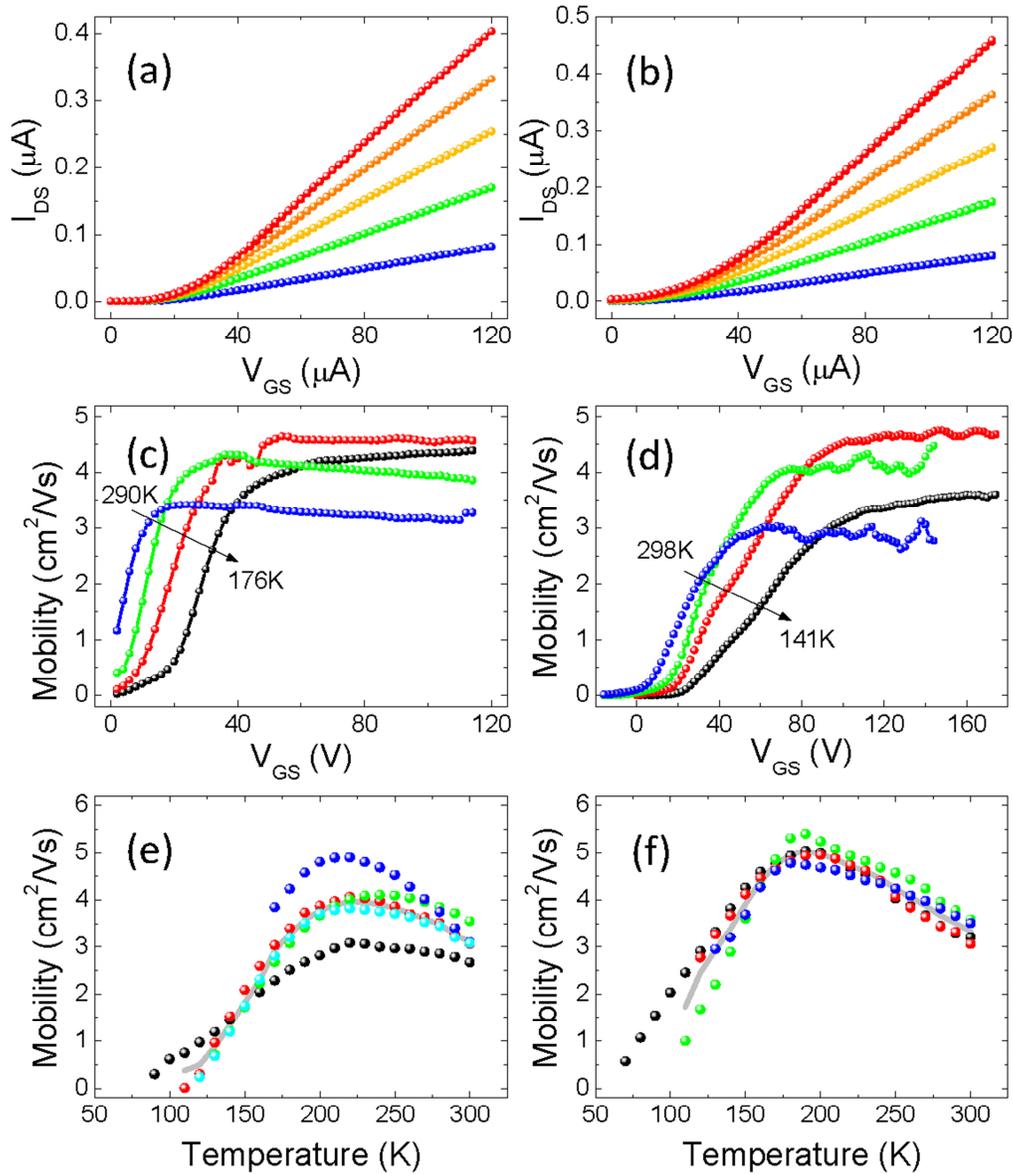

**Figure 2.** Room temperature transfer characteristics of single-crystal air-gap PDI3F (a) and PDI5F (b) OFETs for $V_{DS}$ = 5, 10, 15, 20, 25 V (the channel length is approximately 200 µm). Panels c) and d) show the corresponding gate voltage dependence of the mobility (PDI3F (c) and PDI5F (d)) at different temperatures. Above the threshold region, the mobility does not significantly depend on $V_{GS}$. The temperature-dependence of $\mu$ for several air-gap devices of PDI3F and PDI5F is shown in panels e) and f), from which the high reproducibility of the measurements and the value of $T^*$ can be inferred. The solid lines represent the mobility averaged over all samples.

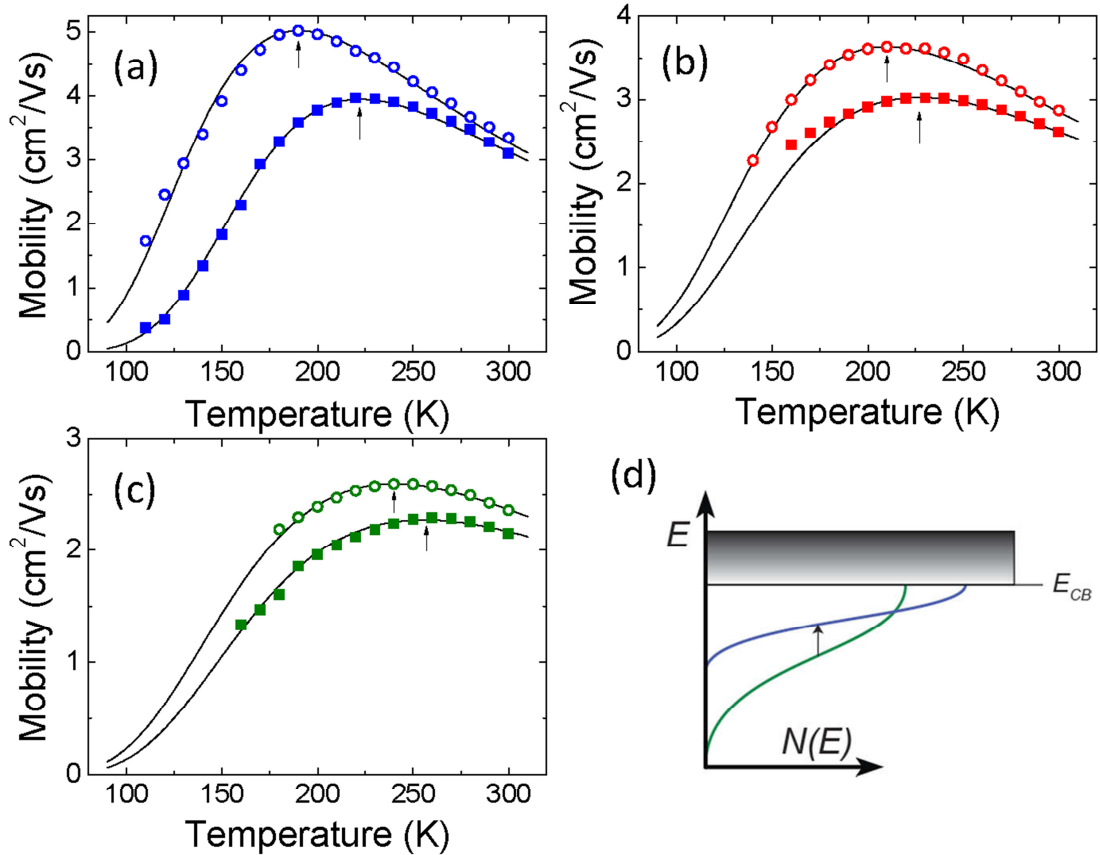

**Figure 3.** Averaged temperature dependent mobility $\bar{\mu}(T)$ of PDI3F (full squares) and PDI5F devices (open circles) with vacuum (a), Cytop (b) and PMMA (c) gate dielectrics. Panel (a) shows that in the intrinsic regime – i.e., in the range where $\mu$ increases upon lowering $T$ – $\mu$ nearly coincides for both molecules, while deviations in $\mu(T)$ become larger in the activated regime below $T^*$ (indicated by the arrows). The continuous lines are fits to $\bar{\mu}(T)$ by the model discussed in the main text. The different $T$-ranges for air-gap, Cytop and PMMA devices are due to cracking of crystals on Cytop and PMMA at higher $T$ (due to differences in the thermal expansion coefficient of the substrates and the organic material). Panel (d) schematically illustrates the energy dependent density of states assumed by the model. The two Gaussian curves indicate that, for any given dielectric, the width of the disorder-induced band tail is different for PDI3F and PDI5F.

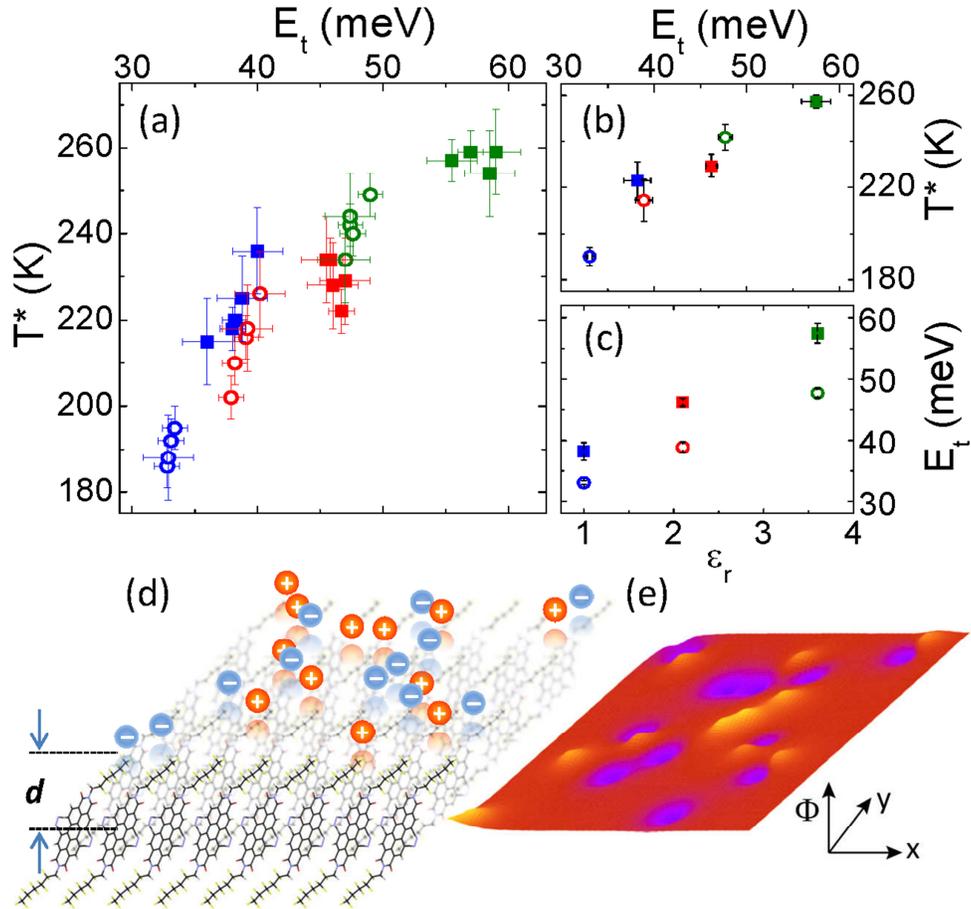

**Figure 4.** (a) Plot of $T^*$ versus $E_t$ extracted from the $\mu(T)$ curves of all FETs investigated on PDI3F (full squares) and PDI5F (open circles) with vacuum (blue), Cytop (red) and PMMA (green) as gate dielectrics. The errors on $E_t$ represent the uncertainties originating from the fitting procedure, while the $T^*$ is mainly determined by the flat maximum in the $\mu(T)$ curves. (b) Relation between $E_t$ and $T^*$, obtained by averaging the data from all individual devices, shown in (a). Panel (c) shows that on each gate dielectric, the value of $E_t$ (averaged over all devices) is systematically smaller for PDI5F (open circles) as compared to PDI3F devices (filled squares): extending the chains in PDI3F by two $C_2F_4$ units decreases $E_t$ by an amount comparable to a change in gate dielectric. Panels (d) and (e) schematically illustrate the mechanism giving the largest contribution to the disorder experienced by the charge carriers: charged particles unintentionally present at the crystal surface (not to scale; the density of surface charges is approximately $10^{-4}$/molecule) generate fluctuations of the electrostatic potential $\Phi$ in the FET channel (e), which localize charge carriers.

# Supporting Information

## 1. Experimental

*Synthesis of mixture of N, N'-bis(1H, 1H-perfluorohexyl)-1,7&1,6-dibromoperylene-3,4:9,10-bis- (dicarboximide) (PDI5F-Br$_2$):* The mixture of 1,7-dibromoperylene-3,4:9,10-tetracarboxylic dianhydride and 1,6-dibromoperylene-3,4:9,10-tetracarboxylic dianhydride (2.83 g, 5.14 mmol) in 1-methyl-2-pyrrolidinone (NMP) (64 ml) was sonicated for 30 min before successive addition of a solution of 1$H$, 1$H$-perfluorohexylamine (5 g, 16.7 mmol) in NMP (43 ml) and acetic acid (2.0 ml). The mixture was stirred at 90 $^o$C for 41 h. After cooling down, the mixture was concentrated to ~20 ml and then precipitated into methanol (100 ml). The precipitates were collected by filtration, washed by methanol and dried to give a black solid. Further purification by column chromatography with chloroform as the eluent gave a red solid (4.51 g, yield 79%). $^1$H NMR (CDCl$_3$, 500 MHz): δ 9.57-9.52 (m, 2H), 9.03-8.98 (m, 2H), 8.81-8.76 (m, 2H), 5.11-5.00 (m, 4H). Elemental Analysis for C$_{36}$H$_{10}$Br$_2$F$_{22}$N$_2$O$_4$: Calcd. C 38.88, H 0.91, N 2.52; Found: C 38.98, H 1.01, N 2.60.

*Synthesis of mixture of N, N'-bis(1H, 1H-perfluorohexyl)- 1,7&1,6-dicyanoperylene-3,4:9,10-bis- (dicarboximide) (PDI5F-CN$_2$):* CuCN (2.00 g, 22.4 mmol) was added to the mixture of the above dibromoperylene (6.18 g, 5.56 mmol) and DMF (282 ml). The mixture was heated to 150°C and stirred for 24 hr. After cooling down, filtered and washed with MeOH several times. The crude brown solid (5.41 g) was purified by vacuum sublimation twice at 310 $^o$C under $5 \times 10^{-5}$ torr to give a red solid (1.05 g, yield 18%). The solubility is too low for NMR. Elemental Analysis for C$_{38}$H$_{10}$F$_{22}$N$_4$O$_4$: Calcd. C 45.44, H 1.00, N 5.58; Found: C 45.35, H 1.00, N 5.45.

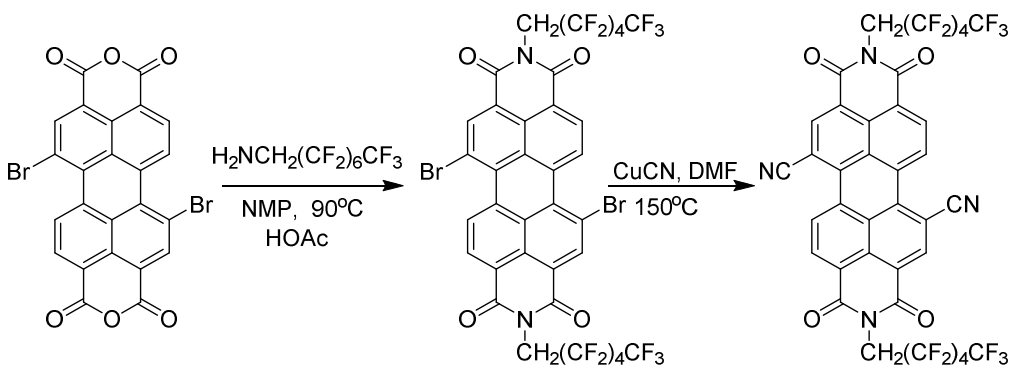

*Crystal growth:*

Single crystals of PDI3F and PDI5F were grown by vapor phase transport in a flow of ultrapure argon gas at a sublimation temperature of approximately 290 °C. Most crystals were platelet-like, with typical dimensions of 100-200 µm width and ~ 300-500 µm length, were highly uniform and showed smooth surfaces. Crystals used for electrical measurements were re-crystallized in order to minimize impurities.

*Device fabrication:*

Polydimethylsiloxane (PDMS) air-gap stamps were prepared according to previously reported procedures[S1] by pouring a 10:1 w/w solution of base:curing agent onto a pre-fabricated SU-8/25 photoresist mould (10 µm thickness) with subsequent baking in an oven for 24 hours at 100°C. 3/20nm of Ti/Au was evaporated in an e-beam evaporator ($p$ = 5·10$^{-8}$ mbar) at 0.1 nm/s to form (electrically isolated) source, drain and gate contacts as well as voltage probes. The field-effect transistors were completed by manually laminating single crystals onto the pre-fabricated contacts using the hair of a fine brush. Cytop and PMMA substrates were fabricated as in Ref. [S2] by spin-casting the dissolved polymers onto heavily doped Si wafers with a 290 nm layer of thermally grown $SiO_2$ and subsequent baking on a hot plate. The thickness of the baked layers (typically around 50 nm for Cytop and 200 nm for PMMA) were measured with a surface profiler and showed high homogeneity. The capacitances of the $SiO_2$/polymer dielectrics were determined in a parallel-plate geometry using an Agilent 4284A LCR meter and coincided with the calculated values. 15 nm thick Au contacts were evaporated onto the substrates using SU8 shadow masks. In all devices, the channel length varied between 160 and 240 µm, and the channel width was determined by the width of the crystal used (typically ~100 µm)

*Electrical measurements*:

Electrical measurements were performed using an Agilent E5270B parameter analyzer in the vacuum chamber ($p$ = 5·10$^{-7}$ mbar) of a helium flow cryostat. Special attention was given to the variable temperature measurements in order to ensure temperature stability and minimize the probability of crystal cracking at high temperature (on PMMA and Cytop devices, crystal cracking at low temperature appears to be unavoidable): the devices were cooled at a constant slow rate of 0.5 K/min and the temperature was stabilized for 10 minutes prior to every measurement. The set temperature was varied in steps of 10 K. The drain-source bias used to determine the four-terminal linear FET mobility was $V_{DS}$ = 5 V.

*Temperature-dependence of the mobility for Cytop and PMMA FETs*

As we repeatedly mentioned in the main text, having an excellent reproducibility of the temperature dependence of the mobility in different devices is essential to compare the behavior of single-crystal based on PDI3F and PDI5F. In the main text we have shown (Figure 2) the reproducibility of the *μ(T)* curves measured on air-gap devices, for both of these molecules. Here, in Figure S1, we show data of the linear field-effect mobility *μ(T)* obtained from four-terminal measurements on several PDI3F single crystal FETs on Cytop (Figure S1a) and PMMA (Figure S1b) gate insulators, as well as the corresponding data for PDI5F devices (Figure S1c,d). The solid gray lines are the average *T*-dependent mobility $\bar{\mu}(T)$ shown in Figure 3b,c of the main text. The data clearly indicate that, as in the case of air-gap devices, PDI3F and PDI5F devices with Cytop and PMMA gate dielectrics also show extremely high reproducibility of the *μ(T)* curves, as compared to most organic FETs reported in the literature.

## 2. Fitting of the *μ(T)* data

In order to fit the *μ(T)* data, we use a phenomenological mobility edge model where the density of states (DOS) consist of a band in which electrons move with an intrinsic mobility $\mu_0(T)=\alpha \cdot T^{-2}$ ($\alpha$ = 3.5·10$^5$ cm$^2$K$^2$V$^{-1}$s$^{-1}$ is the same in all cases) and of a band tail caused by disorder in the crystal and in the environment, in which electrons are localized. $N_t$ and $E_t$ are the height and width of the Gaussian-like distribution of tail states $N_t(E) = N_t \cdot exp(-\frac{E^2}{2E_t^2})$ below the conduction band (CB), where the DOS is taken to be $N_0$ = 10$^{15}$ cm$^{-2}$eV$^{-1}$ (corresponding to the density of molecules at the surface divided by the bandwidth, which is approximately ~0.5 eV). In practice, $\mu(T)=\mu_0(T) \cdot n_b(E_F)/n_{tot}$ is calculated by first evaluating the position of the Fermi energy $E_F(T)$ for a given density of states numerically and then determining the fraction of occupied CB states $n_b(E_F)/n_{tot}$. The fit to the *μ(T)* data is done by varying the height and width of the trap distribution (see Refs. [S2-S4] for further details on the model).

## 3. Theoretical analysis

In the main text we have considered potential fluctuations originating from charged impurities unintentionally present at the surface of the organic material, as well as from dipoles in the gate insulator. We have shown that such electrostatic effects can explain (at a rather good quantitative level) the observed changes in the disorder parameter $E_t$ extracted from experiments on the whole series of devices studied. Here we provide some details on the theoretical analysis, and discuss other microscopic mechanisms that we considered and found to be incompatible with the experimental observations, or that do not to play a relevant role.

### 3.1 Microscopic estimates of potential fluctuations

To determine the spread of potential fluctuations in the conducting channel of our single-crystal FETs (i.e., the outermost plane of conjugated cores of PDI3F/PDI5F molecules) we assume a random distribution of charged adsorbates located at the surface, i.e. at a distance $d$ from the channel (with $d = c/2$ equal to half the inter-layer distance in the crystal structure). The result [S5] is Eq. (1) of the main text. It depends on the inverse screening length $q_s$, which characterizes screening by mobile charges in the channel. For the estimates of $\Delta$ made in the main text, we assume Debye screening, which is appropriate for non-degenerate charge carriers, and gives $q_s = (2\pi e^2/\tilde{\varepsilon})\, n_{2D}/k_B T$ (for the definition of the parameters entering the previous equations see the main text). The value that we use in our calculations, $q_s = 0.02$ Å$^{-1}$, corresponds to taking a typical carrier density of $n_{2D} = 10^{11}$ cm$^{-2}$ at $T = 150$ K. Note that, although we have described explicitly the case of charged impurities located at the surface of the organic crystal, analogous results are obtained if the charged impurities are present in the bulk of the organic crystal.

## 4. Other mechanisms

As discussed in the main text, the dependence of extrinsic disorder on the ligand length can be explained satisfactorily (both qualitatively and quantitatively) in terms of electrostatic potential fluctuations generated from charges (or dipoles) located outside the FET channel. To reach this conclusion, we have considered theoretically several other different mechanisms that can lead to variations in the carrier mobility when changing the length of the ligands. These mechanisms can be classified in two main categories: (I) mechanisms which directly modify the amount of disorder felt by the mobile electrons and (II) mechanisms which modify the hopping integrals, resulting in a change of the mobility-vs-temperature, curves even though the absolute strength of disorder is nominally unchanged.

In this context, we can invoke results of theoretical calculations of Ref. [S6] showing that mechanisms of type (I) are compatible with the experimental observations, while mechanisms of type (II) are not. Specifically, calculations show that reducing the amount of disorder – i.e., a type (I) mechanism – has the effect of increasing the mobility at low temperature, in the range where the mobility is limited by disorder, while leaving the mobility unchanged at higher temperature, in the $T$ range where the intrinsic "band-like" behavior is observed (see Figure 2 of Ref. [S6]). With reducing disorder, the transition from these low- and high-temperature regimes is accompanied by a reduction of the temperature $T^*$ at which the mobility is maximum, as shown in Figure S2a. This evolution of the $\mu(T)$ curves with reducing disorder reproduces precisely what is seen in the experiments when comparing PDI3F and PDI5F single-crystal FETs (see Figure 3 of the main text). Using the same theoretical calculations of Ref. [S6] it can be shown that modifying the hopping integrals (type (II) mechanism) results in a qualitatively different evolution, with an overall increase/decrease of the whole mobility curve, which leaves $T^*$ unchanged (see Figure S2b). Changes in the hopping integral, therefore, do not reproduce the experimental observations presented in this work. In other words, these considerations indicate that the experimental evolution of the $\mu(T)$ curves when passing from PDI3F to PDI5F is consistent with an effective decrease of strength of extrinsic disorder experienced by the carriers residing in the FET channel (the outermost plane formed by the conjugated molecular cores of the molecules) and cannot be explained with differences in hopping integrals in the two cases.

Within the mechanisms of type (I), short-ranged impurities can be excluded right from the start, as they are incompatible with the gradual dependence with respect to the impurity-channel distance $d$, which is found in the experiments. Moving to long-ranged impurities, both charges and dipoles adsorbed on the surface of the material are a possibility. For the case of air-gap devices, if the external potential were generated by dipoles, the experimental observations could only be reproduced by assuming an unrealistically large surface concentration, corresponding to one impurity dipole per organic molecule. Such a large concentration can be excluded on experimental basis for air-gap devices. Clearly, for devices on PMMA or Cytop, dipoles are present at such large concentrations – and taken into account

– since they originate from the monomers present in the polymeric gate dielectric, as we have discussed in detail in the main text.

Finally, we have considered two other potentially relevant physical phenomena. First, changing the ligand length would alter the dielectric polarizability. Since the polarizable part of the molecule is the conjugated core, increasing the ligand length would result in a decrease of the dielectric constant of the organic material in the relevant direction perpendicular to the planes, which would affect the electrostatic potential fluctuations experienced by charge carriers in the channel (since a larger dielectric constant suppresses the potential fluctuations more). According to this mechanism, owing to its longer ligand length, PDI5F should possess a smaller dielectric constant resulting in larger potential fluctuations, as compared to PDI3F, and therefore a stronger disorder. This is the opposite of what we find experimentally, implying that this mechanism is not the most relevant one. Second, we have considered the tendency towards self-trapping of carriers due to their interaction with the dynamical electronic polarization of the molecules. These polarization effects may further localize carriers, which are already localized by intrinsic and extrinsic disorder. They would lower the energy of the electronic states by an amount which is inversely proportional to their original localization length, thereby effectively causing a deepening of the distribution of sates in the tail of the band. However, theoretically such a mechanism only plays a role if the polarization dynamics is sufficiently slow, since, as demonstrated in Ref. [S7], electrical polarization with a fast dynamics only causes a renormalization of the hopping integrals. As we discuss above, however, a hopping integral renormalization pertains to type (II) mechanisms, and cannot explain our experimental observations.

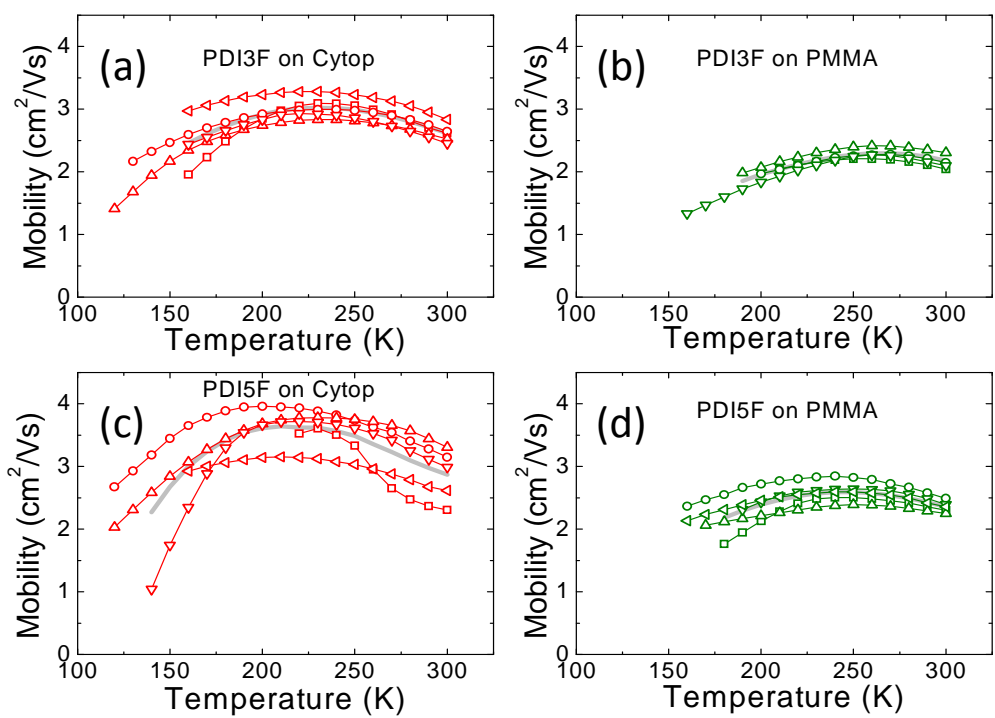

**Figure S1.** Temperature-dependence of the four-terminal linear field-effect mobility $\mu(T)$ for several PDI3F single crystal FETs with Cytop (a) and PMMA (b) gate dielectrics. The corresponding $\mu(T)$ data of PDI5F FETs is shown in panels (c) and (d). The solid lines represent the mobility averaged over all samples. The channel length in all cases was approximately 200 µm.

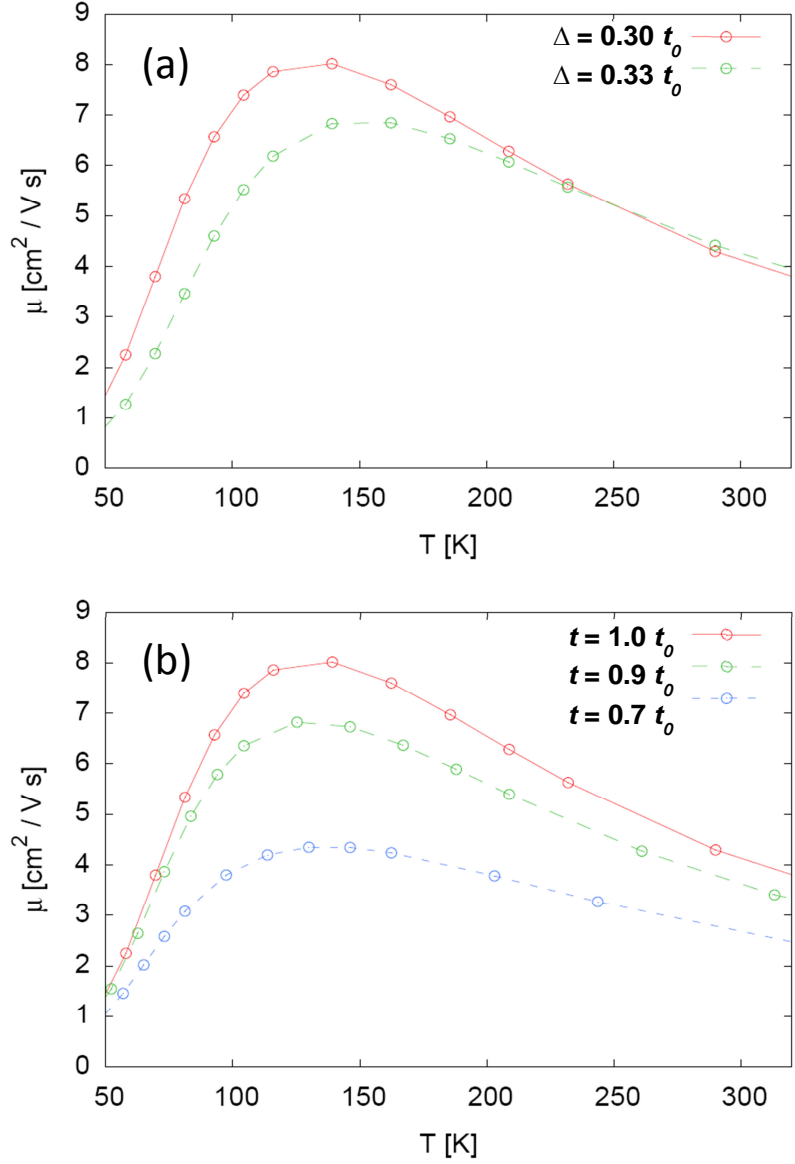

**Figure S2.** Calculations of a microscopic transport model, reported in Ref. [6], showing how changes in disorder energy $\Delta$ (a) or hopping integral $t$ (b) affect differently the behavior of $\mu(T)$. Parameters are the same as in Ref. [6], with $t_0 = 100$ meV. Panel (a) shows how a change of $\Delta$ shifts the position of $T^*$, while leaving the intrinsic mobility regime at high temperature unchanged (the hopping integral is fixed to $t = t_0$). Panel (b) shows how modifying the hopping integral $t$, for a given disorder energy $\Delta$ results in an overall increase/decrease of $\mu(T)$, while $T^*$ remains unchanged (here the extrinsic disorder $\Delta$ is taken to be constant and equal to $0.3\,t_0$)